\documentclass[12pt,twoside]{article}
\usepackage[T2A]{fontenc}
\usepackage[cp1251]{inputenc}
\usepackage[russian]{babel}
\usepackage{floatflt}
\usepackage{CPMJ}
\usepackage{listings}
\usepackage{amssymb}
\usepackage{amsfonts}
\usepackage{amsmath}
\usepackage{amsthm}
\usepackage{graphicx, graphics}
\usepackage{caption, psfrag}
\usepackage{xspace}
\begin{document}

\newpage
\noindent{\footnotesize{\it }\vspace{5mm}

\noindent {сдй 517.984}}\vspace{3mm}

\author{$^1$Muzaffar M. Rahmatullaev and $^2$Bunyod U. Abraev}
%нТНПЛКЕМХЕ ГЮЦКЮБХЪ
\title{A NEW SET OF GIBBS MEASURES FOR THE SOS MODEL ON A CAYLEY TREE}
%нТНПЛКЕМХЕ ЯЯШКЙХ МЮ ЦПЮМР
%\ack{пЮАНРЮ БШОНКМЕМЮ ОПХ ОНДДЕПФЙЕ ЦПЮМРЮ пттх (ОПНЕЙР 07-01-96002-П$\_$СПЮК$\_$Ю).}

%%%%%%%%%%%%%%%%%%%%%%
%
\maketitle              %ЙНЛЮМДЮ ОНЙЮГШБЮЕР РНКЭЙН \author Х \title Х ВХЯРХР ЯВЕРВХЙХ

\thispagestyle{empty}
\vspace{3mm}
\noindent{\small{\it $^1$Institute of mathematics after named V.I.Romanovsky, Tashkent, Uzbekistan \\
$^2$Chirchik state pedagogical institute of Tashkent region, Chirchik, Uzbekistan}}

%\noindent{\small{\it ${}^2$чФМН-сПЮКЭЯЙХИ ЦНЯСДЮПЯРБЕММШИ СМХБЕПЯХРЕР\\ (МЮЖХНМЮКЭМШИ ХЯЯКЕДНБЮРЕКЭЯЙХИ СМХБЕПЯХРЕР), вЕКЪАХМЯЙ}}

\noindent{\small{\it $^1$mrahmatullaev@rambler.ru,  $^2$abrayev89@mail.ru}}

%%%%%%%%%%%%%%%%%%%%%%
% юММНРЮЖХЪ
\begin{abstract}
\noindent {\footnotesize
The phase transition phenomenon is one of the central problems of statistical mechanics. It occurs when the model possesses multiple Gibbs measures. In this paper, we consider a  three-state SOS (solid-on-solid) model  on a Cayley tree. We reduce description of Gibbs measures to solving of a non-linear functional equation, which each solution of the equation corresponds to a Gibbs measure. We give some sufficiency conditions on the existence of multiple Gibbs measures for the model. We give a review of some known (translation-invariant, periodic, non-periodic) Gibbs measures of the model and compare them with our new measures. We show that the Gibbs measures found in the paper differ from the known Gibbs measures, i.e, we show that these measures are new. }
\end{abstract}
\ekeywords{SOS model, Cayley tree,  Gibbs measures, phase transitions.}
%
%%%%%%%%%%%%%%%%%%%%
%йНКНМРХРСКШ
\markboth{}{} \markboth{Muzaffar M. Rahmatullaev and Bunyod U. Abraev
}{\footnotesize A new class of Gibbs measures for three-state SOS model on a Cayley tree}

\section{Introduction}\label{sec1}

In this paper, we consider the three-state SOS model on a Cayley tree. For this model we give a very wide class of new Gibbs measures. One of the most interesting problems in statistical mechanics on a lattice is the phase transition problem, i.e., deciding whether there are many different Gibbs measures associated to a given Hamiltonian  \cite{B1},\cite{YuRB}, \cite{G}, \cite{B}. In this paper, we  find the sufficiency conditions on the existence of at least three Gibbs measures of the model. The same problem for the Ising model is considered in \cite{4}.

It is believed that several of its interesting thermal properties could persist for regular lattices,
for which the exact calculation is far intractable. Here we mention that in the 90's a lot of
research papers were devoted to the study of the classical Ising model, with two spin values
$\pm 1,$ on such a Cayley tree \cite{B1}, \cite{G}, \cite{B}, \cite{7}, \cite{S}.

In \cite{ART}, the authors constructed some Gibbs measures (hereinafter referred to as the Gibbs measures obtained by the ART construction) for the Ising model on the Cayley tree. In the papers \cite{RJP}, \cite{RDAN} for the Ising model using the translation-invariant (resp. periodic) Gibbs measures on the Cayley tree of order $k_0$, a new Gibbs measure on the Cayley tree of order $k$ $(k_0<k)$ is constructed, called as $(k_0)$-translation-invariant (resp. $(k_0)$-periodic) Gibbs measure.

In \cite{Abrayev} we consider a SOS model with nearest-neighbour interactions and with three spin values, on a Cayley tree of order $k\geq3.$ For this model, using the well-known translation-invariant Gibbs measures \cite{KRSOS}, some non-translation-invariant Gibbs measures are constructed.

The {\it Cayley tree} $\Gamma^k$ of order $k \geq 1$ is an infinite tree, i.e., a graph
without cycles, such that exactly $k+1$ edges originate from each vertex. Let $\Gamma^k=(V,L)$ where
$V$ is the set of vertices and $L$ is the set of edges. Two vertices $x$ and $y$ are called {\it nearest neighbors}
if there exists an edge $l \in L$ connecting them and we denote $l=\langle x,y\rangle$. A collection of nearest neighbor pairs
$\langle x,x_1\rangle, \langle x_1,x_2\rangle, \dots, \langle x_{d-1},y\rangle $ is called a {\it path} from $x$ to $y$.

On this tree, there is a natural distance to be denoted by $d(x,y)$, being the number of nearest neighbor pairs  of the minimal path between  the vertices $x$ and $y$.

For a fixed $x^0\in V$, the root,
let
\begin{equation*}
W_n=\{x\in V :\, d(x,x^0)=n\}, \ \
V_n=\{x\in V :\, d(x,x^0)\leq n\}
\end{equation*}
be respectively the sphere and the ball
of radius $n$ with center at $x^0$,  and for
$ x\in W_n$ let
$$S(x)=\{y\in W_{n+1} :  d(y,x)=1\}$$
be the set of direct descendants (successors) of $x$.

We assume that the spins take values in the set $\Phi:=\{0,1,2, ..., m \}, \, m\geq1$. A configuration $\sigma$ on $A\subseteq V$ is defined as the function $x\in A\mapsto\sigma_A(x)\in\Phi$. The set of all configurations coincides with $\Omega_A=\Phi^A$.
Put $\Omega=\Omega_V$ and $\sigma=\sigma_V$.

The SOS model is defined by the formal Hamiltonian
$$
H(\sigma)=-J \sum_{\langle x,y\rangle\in L} \mid\sigma(x)-\sigma(y)\mid,
$$
where $J\in \mathbb{R}$ is the coupling constant .

Let $h:x\mapsto h_x=(h_{0,x},h_{1,x}, ..., h_{m,x}) \in \mathbb{R}^{m+1}$ be a vector function of $x\in V\setminus\{x^0\}$. Consider a probability distribution $\mu^{(n)}$ on $\Omega_{V_n} $:
$$
\mu^{(n)}(\sigma_n)=Z^{-1}_n \exp(-\beta H(\sigma_n)+\sum_{x\in W_n} h_{\sigma(x),x}), \eqno(1)
$$
where $\sigma_n\in \Omega_{V_n}$ and

$$Z_n=\sum_{\tilde {\sigma}_n\in \Omega_{V_n}}\exp(-\beta H (\tilde {\sigma}_n)+\sum_{x\in W_n} h_{\tilde{\sigma}(x),x}). $$

A probability distribution $\mu^{(n)}$ is said to be compatible if for any $n\geq1$ and $\sigma_{n-1}\in\Omega_{V_{n-1}},$ we have

$$
\sum_{\omega_n\in \Omega_{W_n}} \mu^{(n)}(\sigma_{n-1}\vee\omega_n)=\mu^{(n-1)} (\sigma_{n-1}),
$$
where $\sigma_{n-1}\vee\omega_n\in\Omega_{V_n}$.

In this case, there is a unique measure $\mu$ on $\Omega$ such that $$\mu(\{\sigma\mid_{V_n}=\sigma_n\}) = \mu^{(n)} (\sigma_n)$$ for all $n$ and $\sigma_n\in\Omega_{V_n}$. Such a measure is called a splitting Gibbs measure (SGM) corresponding to the Hamiltonian $H$ and to the function $x\mapsto h_x,x\neq x_0$ \cite{rs}.

The following statement gives a condition on $h_x$  guaranteeing that the distribution of $\mu^{(n)}(\sigma_n)$ is compatible.

\begin{etheorem}\label{theor1}\cite{rs} \textit{\it The probability distributions $\mu^{(n)}(\sigma_n), n=1, 2,... $ determined by the formula (1) are compatible iff for any $x \in V \setminus \{x^0\}$ the following equation holds:}
$$
h^*_ x =\sum_ {y\in S(x)}F(h^*_ y,m,\theta), \eqno(2)
$$
\textit{where $\theta = e^{J\beta}, \beta = \frac{1}{T}$. Here $h^*_x=(h_{0,x}-h_{m,x},h_{1,x}-h_{m,x},...,h_{m-1,x}-h_{m,x})$  and
$F( \bullet ,m,\theta): \mathbb{R}^m \rightarrow \mathbb{R}^m$ is a vector function, i.e.
}
$$F(h,m,\theta)=(F_0((h,m,\theta),...,F_{m-1}(h,m,\theta))), \ \ \ h=(h_0,h_1,...,h_{m-1}),$$
\textit{such~that}

$$
F_i(h,m,\theta)=\ln\frac{\sum^{m-1}_{j=0}\theta^{\mid i-j\mid}e^{h_j}+\theta^{m-i}}{\sum^{m-1}_{j=0} \theta^{m-j}e^{h_j}+1}, \ \ \ i=0,...,m-1.
$$
\end{etheorem}

Namely, for any boundary condition satisfying the functional equation (2) there exists a unique SGM. In this paper we consider only SGMs and omit the word `splitting'.

The paper is organized as follows: The results are given in Section \ref{Res} \, Section \ref{Res1} contains a review of some known Gibbs measures of the SOS model on a Cayley tree and their comparison with the new measures of this paper.

\section{Results}\label{Res}

From now on, we restrict ourselves to the case $m = 2.$ The set of vertices of a Cayley tree of order $k$ is denoted by $V^k.$ For $x\in V$ by $S_{k_0}(x)$ we denote arbitrary $k_0$ vertices of the set $S(x).$

If on the vertex $x$ we have $h_x=\bar{h}$, then to the vertex $S_a(x)$ we put the value $h_x=\bar{h}$, to the other vertices $S_b(x)$ we put the value $h_x=\bar{l}.$ If on the vertex $x$ we have $h_x=\bar{l}$, then on the vertex $S_c(x)$ we put the value $h_x=\bar{h}$, on the remaining vertices $S_d(x)$ we put $h_x=\bar{l}$. Note that $a+b=k, \ c+d=k.$ Using this construction for $\bar{h}=(h_1,h_2)$ and $\bar{l}=(l_1,l_2)$ due to Theorem \ref{theor1} we have
$$
\left\{
\begin{array}{ll}
    \bar{h}=a\,F(\bar{h}, m,\theta)+b\,F(\bar{l}, m, \theta), \\
    \bar{l}=c\,F(\bar{h}, m,\theta)+d\,F(\bar{l}, m, \theta), \\
\end{array}\right. \eqno(3)
$$
or
$$
\left\{%
\begin{array}{ll}
h_1=a\,\ln\frac{\exp{(h_{1})}+\theta\exp{(h_{2})}+\theta^2}{\theta^2\exp{(h_{1})}+\theta
    \exp{(h_{2})}+1}+b\,\ln\frac{\exp{(l_{1})}+\theta\exp{(l_{2})}+\theta^2}{\theta^2\exp{(l_{1})}+\theta
    \exp{(l_{2})}+1}, \\
h_2=a\,\ln\frac{\theta\exp{(h_{1})}+\exp{(h_{2})}+\theta}{\theta^2\exp{(h_{1})}+\theta\exp{(h_{2})}+1}+
b\,\ln\frac{\theta\exp{(l_{1})}+\exp{(l_{2})}+\theta}{\theta^2\exp{(l_{1})}+\theta\exp{(l_{2})}+1},\\
l_1=c\,\ln\frac{\exp{(h_{1})}+\theta\exp{(h_{2})}+\theta^2}{\theta^2\exp{(h_{1})}+\theta
\exp{(h_{2})}+1}+
d\,\ln\frac{\exp{(l_{1})}+\theta\exp{(l_{2})}+\theta^2}{\theta^2\exp{(l_{1})}+\theta
\exp{(l_{2})}+1}, \\
l_2=c\,\ln\frac{\theta\exp{(h_{1})}+\exp{(h_{2})}+\theta}{\theta^2\exp{(h_{1})}+\theta\exp{(h_{2})}+1}+
d\,\ln\frac{\theta\exp{(l_{1})}+\exp{(l_{2})}+\theta}{\theta^2\exp{(l_{1})}+\theta\exp{(l_{2})}+1}.\\
\end{array}%
\right. \eqno(4)$$

Consider the operator $W:\mathbb{R}^4\to \mathbb{R}^4,$ defined by RHS of (4). Note that the system of equations (4) is the fixed-point equation $h=W(h).$ It is obvious that the following set is invariant with respect to the operator $W$:
$$
I=\{h=(h_1,\ h_2,\ l_1, \ l_2)\in \mathbb{R}^4: \ h_1=l_1=0\}.
$$
The system of equations (4) on the invariant set $I$ has the form

$$
\left\{%
\begin{array}{ll}
h_2=a\,\ln\frac{\exp{(h_{2})}+2\theta}{\theta^2+\theta\exp{(h_{2})}+1}+
b\,\ln\frac{\exp{(l_{2})}+2\theta}{\theta^2+\theta\exp{(l_{2})}+1},\\
l_2=c\,\ln\frac{\exp{(h_{2})}+2\theta}{\theta^2+\theta\exp{(h_{2})}+1}+
d\,\ln\frac{\exp{(l_{2})}+2\theta}{\theta^2+\theta\exp{(l_{2})}+1}.\\
\end{array}%
\right.$$
After denoting
$$
f(x)=f(x,\theta):=\ln\frac{\exp{(x)}+2\theta}{\theta^2+\theta\exp{(x)}+1},
\eqno(5)$$
we have:
$$
\left\{%
\begin{array}{ll}
h_2= a\,f(h_2)+b\,f(l_2),\\
l_2= c\,f(h_2)+d\,f(l_2).\\
\end{array}%
\right. \eqno(6)$$

\begin{elemma}\label{L1} \textit{The function $f(x)$ has the following properties:\newline
i)  The function $f(x)$ is increasing for $\theta<1$ and decreasing for $\theta>1.$\newline
ii) The function is bounded:
 $$\min\left\{\ln\frac{2\theta}{\theta^2+1}, \ \ln\frac{1}{\theta}\right\}<f(x)<\max\left\{\ln\frac{2\theta}{\theta^2+1}, \ \ln\frac{1}{\theta}\right\}.$$
iii)  The function is strictly convex for $\theta>1, \ x\in (0;x^*)$  and for $\theta<1, \ x\in (x^*;+\infty)$\newline
(resp. the function is strictly concave for $\theta<1, \ x\in (0;x^*)$ and for $\theta>1, \ x\in (x^*;+\infty))$
where
$$x^*=\frac{1}{2}\,\ln(2\theta^2+2).$$}
\end{elemma}

\emph{Proof.}  The proofs will be given for $\theta>1.$ The remaining case $\theta<1$ can be proved similarly.

\emph{i)} The derivative of the function $f(x)$ with respect to the variable $x$ has the following form:
$$f'(x)=-\frac{\exp{(x)}(\theta^2-1)}{(\exp{(x)}+2\theta)(\theta^2+\theta\exp{(x)}+1)}.$$
It is easy to see that for $\theta>1$  we have $f'(x)<0$, i.e., the function is decreasing.

\emph{ii)} We have
$$\lim_ {{x\rightarrow +\infty}}{f(x)}=\lim _{{x\rightarrow +\infty}}{\ln\frac{\exp{(x)}+2\theta}{\theta^2+\theta\exp{(x)}+1}}=\ln\frac{1}{\theta},$$
$$\lim_{{x\rightarrow-\infty}}{f(x)}=\lim_{{x\rightarrow-\infty}}{\ln\frac{\exp{(x)}+2\theta}{\theta^2+\theta\exp{(x)}+1}}=
\ln\frac{2\theta}{\theta^2+1}.$$ From these, it follows that for $\theta>1$
$$\ln\frac{1}{\theta}<f(x)<\ln\frac{2\theta}{\theta^2+1},~\forall x\in (-\infty,+\infty).$$

\emph{iii)} We have
$$f''(x)=\frac{\theta(\theta^2-1)(\exp(2x)-2\theta^2-2)\exp(x)}{(\exp(x)+2\theta)^2(\theta^2+\theta\exp(x)+1)^2}.$$
It is easy to see that for $\theta>1, \ x\in (0;x^*)$ we have $f''(x)>0$ and  for $\theta>1, \ x\in (x^*;+\infty))$
we have $f''(x)<0,$ where
$$x^*=\frac{1}{2}\,\ln(2\theta^2+2).$$
This completes proof.

We shall study the system of equations (6) for the cases $b\neq0$ and $b=0$ separately.

Let $b\neq0$. In this case, from the first equation of  (6), we obtain the equality $f(l_2)=\frac{1}{b}\big(h_2-af(h_2)\big).$
From this, using the second equation of (6), we have the following:
    $$l_2=\frac{1}{b}\Big(\big(bc-ad\big)f(h_2)+dh_2\Big).$$
As a result, the first equation of (6) can be written as
$$
h_2=a\,f(h_2)+b\,f\Big(\frac{1}{b}\big((bc-ad)f(h_2)+d\,h_2\big)\Big).
\eqno(7)$$
Denote
$$\varphi(h_2):=\frac{1}{b}\,\Big((bc-ad)f(h_2)+d\,h_2\Big),$$ and
$$\psi(h_2):=a\,f(h_2)+b\,f\Big(\frac{1}{b}\Big((bc-ad)f(h_2)+d\,h_2\Big)\Big).$$

The function $\psi(h_2)$ is continuous and bounded for $h_2\in(-\infty,\infty)$. This implies that it has at least a fixed point, say, $h_2^*.$

\begin{etheorem}\label{TH1} Let $b\neq 0.$ If the condition
$$\left| a\,f'(h^*_2)+\Big(\big(b\,c-a\,d\big)f'(h^*_2)+d\,\Big)f'\Big(\frac{1}{b}\Big((b\,c-a\,d)f(h^*_2)+d\,h^*_2\Big)\Big)\right|>1$$
is satisfied, then for the SOS model there exist at least three Gibbs measures, which correspond to the solutions $(h^*_2, \varphi(h^*_2)),$ $(h^1_2, l^1_2),(h^2_2, l^2_2) $ of the system equations (6).
\end{etheorem}

\emph{Proof.} Let $b\ne 0$. Let $h_2^*$ be a fixed point of the function $\psi.$ Then $$\psi'(h^*_2)= a\,f'(h^*_2)+\Big(\big(b\,c-a\,d\big)f'(h^*_2)+d\Big)f'\Big(\frac{1}{b}\Big((b\,c-a\,d)f(h^*_2)+dh^*_2\Big)\Big)$$ and $\psi(h_2)$ is a bounded function of $h_2$. Moreover, if $\mid \psi'(h^*_2)\mid>1$ (i.e. $h^*_2$ is unstable fixed
point of $\psi$)  then there is a sufficiently small neighborhood of $(-\varepsilon+h^*_2,h^*_2+\varepsilon)$  such
that $\psi(h_2)<h_2$, for $h_2\in(-\varepsilon+h^*_2, \ h^*_2)$ and $\psi(h_2)>h_2$, for  $h_2\in(h^*_2,\ h^*_2+\varepsilon).$  For $h_2\in(h^*_2, \ h^*_2+\varepsilon)$ the iterates remain $\psi^n(h_2)$ $>h^*_2,$ monotonically increase and hence converge to a limit, $h^{(1)}_2>h^*_2$ which solves (7). However, $h^{(1)}_2>h^*_2,$ as $h^*_2$ is unstable. For $h\in(-\varepsilon+h^*_2,h^*_2)$ the iterates remain $\psi^n(h_2)<h^*_2,$ monotonically increase. Hence, it converges to a limit, $h^{(2)}_2<h^*_2$ which is solution of (7). However, $h^{(2)}_2<h^*_2,$ as $h^*_2$ is unstable. Hence, we deduce that if $\mid \psi'(h^*_2)\mid>1$ then the system (6) has at least three solutions $(h_i,\varphi(h_i)), \ i=1,2,3.$ Theorem \ref{TH1} is proved.

\begin{eremark}
Note that if we consider the system of equations (6) by similar way but with respect to unknown parameter $l_2$  then we can find another conditions on parameters which ensures the existence of at least three Gibbs measures.
\end{eremark}

Let $b=0.$ It follows from (6) that

$$
\left\{%
\begin{array}{ll}
h_2= k\,f(h_2),\\
l_2= d\,f(l_2)+c\,f(h_2).\\
\end{array}%
\right. \eqno(8)$$

Note that for any positive $\theta,$ the first equation of system  (8) has at least one solution (see \cite{rs}). Let this solution be $h^*.$ Using this solution, from the second equation of system (8) we have
$$
l_2=d\,f(l_2)+\frac{c\,h^*}{k}. \eqno(9)
$$
Then (9) can be rewritten as
$$
 \exp(l_2)=\exp\Big(\frac{c\,h^*}{k}\Big)\,\left(\frac{\exp(l_2)+2\theta}{\theta^2+\theta \exp(l_2)+1}\right)^d.
$$

Let us first address the solvability of the equation (9). Denote
$$\theta<\theta_c\equiv\theta_c(d)=\frac{d-1}{\sqrt{d^2+6\,d+1}}. \eqno(10)$$

Let us also set
$$\zeta=\zeta(\theta):=\frac{1+\theta^2}{2\,\theta^2}, \eqno(11)$$

Clearly, $\zeta(1)=1$ and $\zeta(\theta)>1$ for $\theta<1.$ Furthermore, comparing (10) and (11) observe
that $\zeta(\theta_c)=\big(\frac{d+1}{d-1}\big)^2$ and $\zeta(\theta)>\big(\frac{d+1}{d-1}\big)^2$
for $\theta<\theta_c.$ For $\theta<\theta_c$, denote by $x_{1,2}=x_{1,2}(\theta)$ the
roots of the quadratic equation
$$x^2+\Big[2-(\zeta-1)\,(d-1)\Big]\,x+\zeta=0  \eqno(12) $$
with discriminant
$$D(\zeta,d):=\Big(2-(\zeta-1)(d-1)\Big)^2-4\zeta=(\zeta-1)(d-1)^2\Big(\zeta-\big(\frac{d+1}{d-1}\big)^2\Big). \eqno(13)$$
The solutions of (12) are
$$x_{1}=\frac{(\zeta-1)\,(d-1)-2-\sqrt{D(\zeta,d)}}{2} \quad \mbox{and} \eqno(14)$$
$$x_{2}=\frac{(\zeta-1)\,(d-1)-2+\sqrt{D(\zeta,d)}}{2}.$$
Furthermore, introduce the notation
$$ \eta_{i}=\eta_{i}(\zeta,\theta):=\frac{1}{x_{i}}\Big(\frac{1+x_{i}}{\zeta+x_{i}}\Big)^d,\quad i=1,2, \quad \theta\leq\theta_c, \eqno(15)$$
Of course, $\eta_1(\theta_c)=\eta_2(\theta_c),$ and one can also show that $\eta_1(\theta)<\eta_2(\theta)$ for all $\theta<\theta_c.$ Finally, denote
$$c_{1}^*=c_{1}^*(\theta):=\frac{k}{h^*}\ln\frac{2\,\theta^{d+1}}{\eta_{2}}\quad\mbox{and}\quad c_{2}^*=c_{2}^*(\theta):=\frac{k}{h^*}\ln\frac{2\,\theta^{d+1}}{\eta_{1}}, \quad \theta\leq\theta_c \eqno(16)$$
so that $c_1^*(\theta_c)=c_2^*(\theta_c),$ and $c_1^*(\theta)<c_2^*(\theta)$ for $\theta<\theta_c.$

\begin{elemma}\label{lem1} Let $N(\theta,c)$ denote the number of solutions $l_2>0$ of the equation (9).
Then
$$N(\theta,c)=\left\{%
\begin{array}{lll}
1 \quad \mbox{if} \quad \theta\geq \theta_c \quad \mbox{or} \quad \theta<\theta_c \quad  \mbox{and} \quad  c\notin[c_1^*,c_2^*];\\
2 \quad \mbox{if} \quad \theta<\theta_c  \quad  \mbox{and} \quad  c\in\{c_1^*,c_2^*\};\\
3 \quad \mbox{if} \quad \theta<\theta_c \quad  \mbox{and}  \quad  c\in(c_1^*,c_2^*).\\
\end{array}%
\right.$$

\end{elemma}

\emph{Proof.} By the substitution
$$x=\frac{\exp(l_2)}{2\theta}$$
equation (9) can be represented in the form
$$\eta\,x=g(x), \quad g(x):=\left(\frac{1+x}{\zeta+x}\right)^d, \eqno(17)$$
with the coefficients
$$\eta=\eta(\theta):=\frac{2\,\theta^{d+1}}{\exp(\frac{c\,h^*}{k})}>0, \quad \zeta=\zeta(\theta):=\frac{1+\theta^2}{2\,\theta^2}\geq1. \eqno(18) $$

Equation (17) is well known in the theory of Markov chains on the Cayley tree (see, e.g., \cite{B}, Proposition 10.7), and it is easy to analyse the number of its positive solutions. The case $\zeta=1$ is obvious. Assuming $\zeta>1,$ it is straightforward to check that
$g(x)$ is an increasing function, with $g(0)=\zeta^{-d}< 1$ and $\lim\limits_{x\rightarrow\infty}g(x)=1;$ also, it has one
inflection point $x_0=\frac{1}{2}(\zeta(d-1)-(d+1)),$ such that $g(x)$ is convex for $x<x_0$ and concave for $x>x_0$ (note that $x_0>0$  only  when $\zeta>\frac{d+1}{d-1}).$  Therefore, the equation (17) has at least one and at most three positive solutions. In fact, by fixing $\zeta>0$  and  gradually increasing the slope $\eta>0$ of the ray $y=\eta x$ $(x\geq0),$ it is evident that there are more
than one solutions (i.e. intersections with the graph $y=g(x)$) if and only if the equation $xg'(x)=g(x)$ has at least one solution, each such solution $x = x_*$ corresponding to the line $y=\eta x,$ with $\eta=g'(x_*),$ serving as a tangent to the graph $y=g(x)$ at point
$x=x_*.$ In turn, from (17) we compute
$$g'(x)=d\left(\frac{1+x}{\zeta+x}\right)^{d-1}\frac{\zeta-1}{(\zeta+x)^2}=g(x)\frac{d(\zeta-1)}{(\zeta+x)(1+x)},\eqno(19)$$
and it readily follows that the condition $xg'(x)=g(x)$ transcribes as the quadratic
equation (12), with discriminant $D$ given by (13). Thus, if $D>0,$  that  is, $\zeta>(\frac{d+1}{d-1})^2,$
then the equation (12) has two distinct roots $0<x_1<x_2,$ corresponding to the 'critical' values $\eta_{1,2}=g(x_{1,2})/x_{1,2}$ (see (14) and (15)). Furthermore, using (19) it is easy to see that the function $x\mapsto g(x)/x$ is increasing on the interval $x\in[x_1,x_2];$ hence, $\eta_1<\eta_2.$

To summarize, if $\zeta\leq (\frac{d+1}{d-1})^2$ then the equation (17) has a unique solution, whereas
if $\zeta> (\frac{d+1}{d-1})^2$ then there are one, two or three solutions according as $\eta\in[\eta_1,\eta_2],$ $\eta\in\{\eta_1,\eta_2\}$ or $\eta\in(\eta_1,\eta_2),$ respectively. Adapting these results to equation (9), in view of the second formula in (18) the condition $\zeta(\theta)>(\frac{d+1}{d-1})^2$ is equivalent to $\theta<\theta_c,$ with $\theta_c =\theta_c(k)$ defined in (10). The corresponding critical values $c^*_{1,2}$ of the field parameter $c^*$ are determined by the first formula in (18), that is,
$$c_{1}^*=c_{1}^*(\theta):=\frac{d}{h^*}\ln\frac{2\,\theta^{d+1}}{\eta_{2}}\quad\mbox{and}\quad c_{2}^*=c_{2}^*(\theta):=\frac{d}{h^*}\ln\frac{2\,\theta^{d+1}}{\eta_{1}}$$
leading to formula (16). This completes the proof of lemma \ref{lem1}.

\begin{etheorem}\label{thm2} \textit{ Let $k\geq2$ and $b=0.$ For the SOS model on the Cayley tree  following assertions hold:\newline
 $\bullet$ There is at least 1 Gibbs measure corresponding to solution $(h^*,l_2^1)$ if $\theta\geq\theta_c~\mbox{or}~\theta<\theta_c~\mbox{and}~c\notin[c_1^*,c_2^*],$ \newline
 $\bullet$ There are at least 2 Gibbs measures corresponding to solutions $(h^*,l_2^1)$ and $(h^*,l_2^2)$ if $\theta<\theta_c~\mbox{and}~c\in\{c_1^*,c_2^*\},$\newline
 $\bullet$ There is at least 3 Gibbs measures corresponding to solutions $(h^*,l_2^1),$ $(h^*,l_2^2)$ and $(h^*,l_2^3)$ if $\theta<\theta_c~\mbox{and}~c\in(c_1^*,c_2^*),$ where $\theta_c$ is given in (10) and $c^{*}_{1,2}=c^*_{1,2}(\theta)$ are defined by (16).}
\end{etheorem}

\emph{Proof.} Let $k\geq2$ and $b=0.$ Then the system of equations (6) is reduced to the equation (9). By Theorem \ref{theor1} we know that there is a bijection between Gibbs measures and the solutions of (9). Due to Lemma \ref{lem1}, if $\theta\geq\theta_c$ or $\theta<\theta_c$ and $c\in[c^*_1,c^*_2]$ we know that there exist one Gibbs measure. Since we study the system of equation (6) on the invariant set $I$ we deduce that the model possesses at least one Gibbs measure. Similarly, if $\theta<\theta_c$ and $c\in\{c^*_1,c^*_2\}$ then there is at least two Gibbs measures and if $\theta<\theta_c$ and $c\in(c^*_1,c^*_2)$ then there is at least three Gibbs measures. Theorem \ref{thm2} is proved.

\begin{eremark}\label{rk1} 1) Gibbs measures mentioned Theorem \ref{thm2} differ from Gibbs measures which found in \cite{Obid} because they correspond to solutions of distinct system of equations.\\
2) Note that under some conditions on parameters the first equation of (8) has up to three solutions (see \cite{rs}). Above in the case $b=0$ we use only one of these solutions. If we consider another solution there may be exist at least three Gibbs measures which different of Gibbs measures obtained in Theorem \ref{thm2}.
\end{eremark}

Gibbs measure corresponding to the $h=(h_1,h_2)$ and $l=(l_1,l_2)$ which solution of the system equation (3) we denote by $\mu_{h,l}$.

\section{ Relation of the measures $\mu_{h,l}$ to known ones}\label{Res1}

{\it Translation-invariant Gibbs measures.} Such measures correspond to $h_x\equiv h,$
i.e. constant functions (see e.g. \cite{rs}). These measures are particular cases of our measures mentioned
in Theorem \ref{TH1} which can be obtained for $a=c.$
Thus from system (6) we obtain the following system:
$$
 \left\{%
\begin{array}{ll}
 h_2= a\,f(h_2)+b\,f(l_2),\\
l_2= b\,f(l_2)+a\,f(h_2),\\
\end{array}%
\right.$$
which yields $h_2=l_2$, where $f(h)$ is defined by (5).  As a result, we get:
$$
h=k\, f(h)=\psi(h). \eqno(20)
$$

If $\theta>1$ then due to Lemma \ref{L1} the function $k\, f(h)$ is decreasing, which implies that the equation (20) has only one solution. Thus we consider the case $\theta<1.$
Under assumptions of Theorem \ref{TH1}  we have
$$
 k\,f'(h^*_2)>1.
$$
In this case it was shown in \cite{rs} that the equation (20) has three solutions.

{\it Periodic Gibbs measures.} Let $G_k$ be a free product of $k+1$ cyclic groups of the second order with generators $a_1, a_2,...,a_{k+1},$ respectively.

It is known \cite{Ro} that there exists a one-to-one correspondence between the set of vertices $V$ of the Cayley tree $\Gamma^k$ and the group $G_k.$
Let $\bar{G_k}$ be a normal subgroup of $G_k$.
\begin{edefinition} A set of vectors $h=\{h_x,x\in G_k\}$ is called $\bar{G_k}$-periodic, if $h_{yx}=h_x$
for any $x\in G_k,$ $y\in \bar{G_k}$.  $G_k$-periodic collection of vectors is called translation-invariant.
\end{edefinition}
\begin{edefinition} A measure $\mu$ is called $\bar{G_k}$ -periodic, if it correspond to a $\bar{G_k}$-periodic set of vectors $h$. $G_k$-periodic Gibbs measures are called translation-invariant.
\end{edefinition}
Let $G^{(2)}_k$ be the subgroup in $G_k$ consisting of all words of even length, i.e. $G^{(2)}_ k = \{x\in G_k:
\mbox{the~length~of~word} ~x~\mbox{is~even}\}$. $G^{(2)}_k$-periodic solutions of (4) are studied in \cite{rs}. These solutions are particular case of Theorem \ref{TH1}, which can be obtained for $a=d=0.$ In this case from system (6) we obtain the following system:
$$
\left\{%
\begin{array}{ll}
h_2= k\,f(l_2),\\
l_2= k\,f(h_2),\\
\end{array}%
\right. \eqno(21)$$
where $f(h)$ is defined by (5).
Denote $g(h)=k\,f(h).$ Then we have from (21) we have
$$
\left\{%
\begin{array}{ll}
h_2= g(l_2),\\
l_2= g(h_2).\\
\end{array}%
\right. \eqno(22)$$

Simplifying (22) we obtain
$$
h_2=g(g(h_2))=\psi(h_2).
$$
It is easy to see that if $\theta<1$ then the function $g(h)$ is increasing which implies that the system (22) has solutions only in the form of $h_2=l_2$, which correspond to translation-invariant Gibbs measures. Since we consider periodic Gibbs measures which is non-translation-invariant, we consider the case $\theta>1.$
Let $h_2^*$ be a solution of $h_2=g(h_2)$, i.e., $h_2^*$ be a translation-invariant solution of the system (22). The existence of this solution independently of remaining parameters is shown in \cite{rs}. We can also show that $h_2^*=g(g(h_2^*))=\psi(h_2^*).$ Then under conditions of Theorem \ref{TH1} we have
$$\mid\psi'(h_2^*)\mid=\mid(g'(h_2^*))^2\mid>1$$
which is
$g'(h_2^*)<-1.$
In this case we recover Proposition 4.3 of \cite{rs} in which it is stated that if $g'(h_2^*)<-1$ is satisfied then the system (22) at least three solutions.

{\it Non-translation-invariant Gibbs measures.}
If $a=c+2$ and $d=b+2,$ then the system of equations (6) can be written as follows

$$
\left\{
\begin{array}{ll}
h_2=(c+2)\,f(h_2)+b\,f(l_2),\\
l_2=c\,f(h_2)+(b+2)\,f(l_2),\\
\end{array}\right. \eqno(23)
$$
where $f(h)$ is defined by (5). The system of equations is studied in \cite{Abrayev} under condition $h_2=2\,f(h_2)$ and $l_2=2\,f(l_2).$ If we consider the cases $h_2\neq2\,f(h_2)$ and/or $l_2\neq2\,f(l_2)$ under conditions of Theorem \ref{TH1} then we obtain Gibbs measures which are different from Gibbs measures found in \cite{Abrayev}. It was stated in \cite{Abrayev} that Gibbs measures corresponding to the solutions of the system of equations (23) are non-translation-invariant.

In this paper, we consider the three-state SOS model on the Cayley tree of general order $k$ and we extend the set of Gibbs measures of this model.
We show that our results recover the results of \cite{Abrayev}, \cite{rs} at some values of parameters.

\textbf{Aknowledgments.} The authors are grateful to the referee for a careful reading of this paper and valuable comments.

%\newpage
%\section*{}\addcontentsline{toc}{section}{яОХЯНЙ КХРЕПЮРСПШ}

\renewcommand{\refname}{References}

{\small
\noindent{\it }

\noindent{\it }}

%\phantom{MMM}\hspace{-30pt}
%\rule{163mm}{1pt}
%\vspace{-42pt}

%\phantom{MMM}\hspace{-30pt}
%\rule{163mm}{1pt}
%\vspace{0mm}

\renewcommand{\refname}{яОХЯНЙ КХРЕПЮРСПШ}

\end{document}